\documentclass[preprint]{aastex62}
\usepackage{natbib}
\submitjournal{ApJ}

\accepted{2 July 2021}

\begin{document}

\title{CHANDRA X-RAY OBSERVATIONS OF V830 TAU: A T TAURI STAR HOSTING AN EVANESCENT PLANET  }

\correspondingauthor{Stephen L. Skinner}
\email{stephen.skinner@colorado.edu}

\author{Stephen L. Skinner}
\affiliation{Center for Astrophysics and
Space Astronomy (CASA), Univ. of Colorado,
Boulder, CO, USA 80309-0389}

\author{Manuel  G\"{u}del}
\affiliation{Dept. of Astrophysics, Univ. of Vienna,
T\"{u}rkenschanzstr. 17,  A-1180 Vienna, Austria}

\newcommand{\ltsimeq}{\raisebox{-0.6ex}{$\,\stackrel{\raisebox{-.2ex}%
{$\textstyle<$}}{\sim}\,$}}
\newcommand{\gtsimeq}{\raisebox{-0.6ex}{$\,\stackrel{\raisebox{-.2ex}%
{$\textstyle>$}}{\sim}\,$}}
\begin{abstract}
\small{A radial velocity study by Donati et al. (2016) reported the detection
of a close-in giant planet in a 4.93 d orbit around the $\sim$2 Myr old weak-lined 
T Tauri star V830 Tau. 
Because of the stringent timescale constraints that a very young host star like 
V830 Tau  would place on hot Jupiter formation models and inward migration mechanisms,
independent confirmation of the planet's existence is needed but so far has
not been obtained.
We present new {\em Chandra} X-ray observations of V830 Tau.
The {\em Chandra} observations in combination with previous {\em XMM-Newton} 
observations reveal strong variable  X-ray emission with an X-ray luminosity spanning the 
range log L$_{x}$ = 30.10 - 30.87 ergs s$^{-1}$.
{\em Chandra} High Energy Transmission Grating (HETG) spectra show emission lines
formed over a range of plasma temperatures from $\sim$4 MK (Ne IX) to  $\sim$16 MK (S XV).
At the separation of the reported planet (0.057 au) the  X-ray flux is 
$\sim$10$^{6}$ - 10$^{7}$ times greater than the Sun's X-ray flux at Jupiter.
We provide  estimates of the X-ray ionization and atmospheric heating rates at
the planet's separation and identify areas of uncertainty that will need to 
be addressed in any future atmospheric models.
}
\end{abstract}
\keywords{stars:individual (V830 Tau) --- stars:pre-main-sequence --- X-rays: stars}
\section{Introduction}
\noindent {\em Hot Jupiters Around Solar-Like Stars}:
The discovery of  a Jupiter-mass planet in a 4.2 d orbit around the $\sim$6 Gyr 
old solar-like star 51 Peg by Mayor \& Queloz (1995) was not anticipated given 
the absence of close-in giant planets in our own solar system.
As a result of ongoing searches for new exoplanets, many such massive
exoplanets in short-period orbits have now been discovered, mostly around older stars.
Such ``hot Jupiters'' typically have masses M$_{p}$ $\geq$ 0.25 M$_{J}$ 
(M$_{J}$ is Jupiter's mass) and orbital 
periods P$_{orb}$ $\ltsimeq$ 10 d, as reviewed by Dawson \& Johnson (2018).
The mechanism(s) by which such massive planets form and end up so close to the host
star is still debated.  Some models posit {\em in situ} formation at close separation
but a more prevalent idea is that they form  at several au beyond the frost line and
migrate inward. The planet's inward migration and ultimate destruction by death-spiral
into the star or tidal breakup is blocked, likely a result of star-planet-disk
interactions involving  angular momentum exchange,
magnetic fields and tidal effects (Strugarek et al. 2017; Dawson \& Johnson 2018).

Recent observational studies are providing tantalizing evidence for the 
presence of hot Jupiters orbiting much younger solar-mass pre-main sequence 
T Tauri stars (TTS). The reported detections are based on radial velocity (RV)
studies which are challenging since they require the faint periodic planetary RV 
signal to be recovered in the presence of  contamination from strong stellar magnetic
activity. Of primary interest here is the $\sim$2 Myr old weak-lined TTS 
V830 Tau (Table 1) for which a massive planet has been reported in a
4.93 d orbit (Donati et al. 2015, 2016, 2017). A hot Jupiter has also
been reported orbiting the $\sim$17 Myr old weak-lined TTS TAP 26 (V1069 Tau) 
with a period of 9 - 11 d and best-fit value P$_{orb}$ = 10.79 d (Yu et al. 2017).
Planets have also been reported around other TTS but at much larger separations
than for hot Jupiters. Perhaps the most compelling example is the detection of one 
or more directly imaged formative planets around the classical TTS LkCa 15 at 
separations $a$ $\approx$ 15 - 20 au (Kraus \& Ireland 2012; Sallum et al. 2015). 
LkCa 15 is a strong X-ray source but the planets are well-shielded from high-energy 
stellar radiation by intervening disk gas (Skinner \& G\"{u}del 2013; 2017).   

Hot Jupiters orbiting TTS  are crucial to developing an understanding of how
the exoplanets form and obviously place tight constraints on the timescale
for any inward migration. They are also excellent laboratories for studying
the effects of stellar X-ray and EUV (XUV) heating on the planet's atmosphere
and photoevaporative mass-loss. The XUV luminosity of TTS is enhanced by
orders of magnitude relative to their older main sequence counterparts.
As a result of intense XUV irradiation, mass-loss from hot Jupiters 
orbiting TTS must take
radiative heating and cooling into account 
(Murray-Clay, Chiang, \& Murray  2009; Owen \& Jackson 2012).

\begin{deluxetable}{ccccccccl}
\tablewidth{0pt}
\tablecaption{V830 Tau: Stellar Properties}
\tablehead{
           \colhead{Sp. Type}           &
           \colhead{Age}                &
           \colhead{M$_{*}$}            &
           \colhead{R$_{*}$}            &           
           \colhead{T$_{\rm eff}$\tablenotemark{a}}   &
           \colhead{L$_{*}$}            &
           \colhead{P$_{rot}$}          &
           \colhead{A$_{\rm V}$}        &
           \colhead{distance}            \\
           \colhead{}                   &
           \colhead{(Myr)}              &
           \colhead{(M$_{\odot}$)}      &
           \colhead{(R$_{\odot}$)}      &
           \colhead{(K)}                &
           \colhead{(L$_{\odot}$)}      &
           \colhead{(d)}                &
           \colhead{(mag)}              &
           \colhead{(pc)}               
                                  }
\startdata
 K7 (1,2)  & $\approx$2.2 (2)     & 1.00$\pm$0.05 (2)  & 2.0$\pm$0.2 (2)  & $\approx$4090 (1,2) & 1.2$\pm$0.3 (2) & 2.741 (3) & 0.4 (1) & 130.4$\pm$0.3 (4)      \\
\enddata
\tablenotetext{a}{The quoted value is the average of 3930 K for type K7.5 (ref. 1) and 4250 K for type K7 (ref. 2).}
\tablerefs{(1) Herczeg \& Hillenbrand (2014) (2) Donati et al. (2015) 
            (3) Grankin et al. (2008)   (4) {\em Gaia} EDR3}
\end{deluxetable}


\noindent {\em V830 Tau and its Evanescent Planet: }
Extended spectropolarimetric and photometric observations of the 
weak-lined TTS V830 Tau in late 2015 and early 2016 revealed   
periodic radial velocity variations that were attributed to a 
close-in giant planet V830 Tau b (Donati et al. 2015, 2016, 2017).
The observations were obtained using the 3.6-m Canada France Hawaii Telescope (CFHT)
and 8-m Gemini North Telescope with the ESPaDOnS high-resolution spectropolarimeter
and a similar spectrometer on the 2-m T\'{e}lescope Bernard Lyot. 
The authors reported that strong magnetic activity was observed and 
filtered out in order to characterize the planet's RV signature.
The orbital period based on all data was determined to be 
P$_{orb}$ = 4.927 $\pm$ 0.008 d and a separation 
of $a$ = 0.057 $\pm$ 0.001 au, about seven times smaller than 
Mercury's semi-major axis. The lower limit on the planet's mass 
is M$_{p}$sin($i$) = 0.57 M$_{J}$, 
or  M$_{p}$ = 0.70 $\pm$ 0.12 M$_{J}$ assumimg  $i$ = 55$^{\circ}$. In the 
absence of a photometric transit detection, the planet's radius is not known.
At an age $\sim$2 Myr, a giant planet orbiting 
V830 Tau  would make it the youngest star so far known to host a hot Jupiter.

An  attempt to confirm the radial velocity detection of V830 Tau b was 
undertaken by Damasso et al. (2020). Their independent study was carried out 
between 2017 - 2020 using the HARPS-N echelle spectrograph mounted on the 
3.6-m Telescope Nazionale Galileo (TNG). Somewhat surprisingly, they did not confirm 
the suspected 4.93 d radial velocity variations. However, they note that their 
non-detection does not conclusively rule out the existence of V830 Tau b since 
an increased level of stellar magnetic activity during their observing period 
would make it more difficult to recover the faint planetary signal. 
Nevertheless it is clear that further observational monitoring of V830 Tau
will be needed to verify the planet's existence.

Our interest in this system arises from the strong variable X-ray emission 
of V830 Tau which was detected in a 2005 {\em XMM-Newton} observation with 
the star captured far off-axis (Franciosini et al. 2007). Their model fits
of  {\em XMM-Newton} spectra revealed high plasma temperatures 
T $\gtsimeq$ 33 MK, as commonly found in magnetically-active late-type
coronal X-ray sources. The star's surface magnetic field was mapped 
by Donati et al. (2017) who also detected optical flares. Other signs of
magnetic activity include highly-variable nonthermal radio emission (Bower et al. 2016).  
The planet's close separation raises the
possibility that its orbital motion could perturb the star's
magnetic field structure and induce flares through tidal effects or 
star-planet magnetic interaction. Although a study of mostly
main-sequence G-M exoplanet host stars by  Poppenhaeger et al. (2010)
found no conclusive evidence for effects on coronal X-ray emission 
(as measured by L$_{x}$/L$_{bol}$) due to star-planet interactions, 
V830 Tau is much younger and T Tauri stars tend to be much more 
magnetically-active and X-ray luminous than older late-type 
main-sequence stars.

Keck interferometer JHK observations of V830 Tau show no significant
near-IR excess (Akeson et al. 2005). Remarkably, within a timespan
of $\sim$2 Myr the inner disk has  been cleared, leaving any 
close-in planet directly exposed to harsh XUV radiation from the star. 
But it is noteworthy that the Keck interferometer observations
weakly resolved V830 Tau at K-band. Several possible explanations
were proposed by Akeson et al. including an unknown companion in
the 50 mas field-of-view or an extended component of stellar
scattered light. This raises the interesting question of whether 
a close-in giant planet might have affected the K-band interferometry.

We present here new {\em Chandra} observations of V830 Tau which 
provide the information on its X-ray properties needed to model 
planet irradiation effects. We provide estimates of X-ray
ionization and heating rates and identify areas of uncertainty
that will need to be addressed in any future models of the 
planet's atmosphere.

\section{X-ray Observations} 

As summarized in Table 2, {\em Chandra}  observed V830 Tau 
in four observations acquired in November 2018 using the 
Advanced CCD Imaging Spectrometer (ACIS-S) and  High-Energy Transmission 
Grating (HETG) spectrometer. The total observing time  was split into
four observations of roughly equal duration as a result of {\em Chandra}'s
operational and thermal constraints. 

Data were reduced using {\em Chandra} Interactive Analysis 
of Observations (CIAO vers. 4.11) software in combination with
CALDB  vers. 4.8.2 calibration data\footnote{Further information on
CIAO and CALDB can be found at https://cxc.cfa.harvard.edu.}. 
Separate spectra and 
X-ray light curves were generated for each observation
in order to search for source variability, and variability
was detected. X-ray spectra and associated response matrix 
files (RMF) and  auxiliary response files (ARF) files were 
extracted  using CIAO {\em specextract}. Energy-filtered light curves
were produced using CIAO {\em dmextract}. 
Undispersed (zero-order) spectra and light curves were extracted from a circular region
of radius 1$''$.5 centered on the source peak.
Background and pileup were negligible. Spectra were analyzed using 
XSPEC vers. 12.10.1 and  CIAO {\em Sherpa} spectral analysis tools.

\begin{deluxetable}{lllll}
\tabletypesize{\small}
\tablewidth{0pt}
\tablecaption{Summary of V830 Tau Chandra Observations}
\tablehead{
\colhead{Parameter} &
\colhead{}          &
\colhead{}          &
\colhead{}          &
\colhead{}          \\
}
\startdata
ObsId                       & 21166              & 21962                 & 21963            & 21964                \\
Start Date (2018)/Time (TT) &  Nov. 15/10:53     & Nov. 16/15:41         &  Nov. 17/03:31   &  Nov. 18/01:09       \\
Stop  Date (2018)/Time (TT) &  Nov. 15/17:59     & Nov. 16/22:41         &  Nov. 17/10:27   &  Nov. 18/07:47       \\
Elapsed Time (s)            &   25,568           & 25,214                &  24,986          &  23,879              \\
Livetime (s)\tablenotemark{b} & 22,949           & 22,789                &  22,792          &  21,715                \\
\enddata
\tablenotetext{a}{Data were obtained using ACIS-S/HETG in faint timed event mode,
a frame time of 3.1 s, and CCD chips S1-S5 enabled. 
The nominal roll angle for all observations
was 107.16$^{\circ}$. The average X-ray centroid position of V830 Tau from all four
observations was (J2000) 
R.A. = 04$^{\rm h}$33$^{\rm m}$10$^{\rm s}$.02, decl. = $+$24$^{\circ}$33$'$43$''$.0 which is
offset by only 0.29$''$ from the {\em Gaia} EDR3 position
R.A. = 04$^{\rm h}$33$^{\rm m}$10$^{\rm s}$.04, decl. = $+$24$^{\circ}$33$'$42$''$.9. }
\tablenotetext{b}{Livetime corresponds to the time during which source data were being collected 
and excludes operational and instrumental overheads such as CCD readout times.}
\end{deluxetable}



\begin{deluxetable}{llccclcl}
\tablewidth{0pt}
\tablecaption{Summary of V830 Tau X-ray Properties (ACIS-S 0-order)}
\tablehead{
           \colhead{ObsId}              &
           \colhead{Rate}               &
           \colhead{Counts}             &
           \colhead{Hardness}           &
           \colhead{E$_{50}$}           &
           \colhead{kT$_{wgtd}$}        &
           \colhead{F$_{x,abs}$}        &
           \colhead{log L$_{x}$}        \\
           \colhead{}                   &
           \colhead{(c ks$^{-1}$)}      &
           \colhead{(c)}                &
           \colhead{Ratio}              &
           \colhead{(keV)}              &
           \colhead{(keV)}              &
           \colhead{(ergs cm$^{-2}$ s$^{-1}$)}  &
           \colhead{(ergs s$^{-1}$)}     }
\startdata
21166  & 5.71$\pm$1.65    & 134 & 0.28     & 1.28 & 1.10\tablenotemark{b} & 4.70$\pm$0.44e-13\tablenotemark{b}  & 30.10\tablenotemark{b}    \\
21962  & 6.00$\pm$2.31    & 141 & 0.35     & 1.63 & 1.84                  & 5.51$\pm$0.53e-13                   & 30.16 \\
21963\tablenotemark{c} & 9.08$\pm$6.70v    & 212 & 0.36   & 1.61 & 1.61   & 7.47$\pm$0.59e-13\tablenotemark{d}  & 30.29 \\
21964  & 11.8$\pm$3.3     & 270 & 0.47     & 1.90 & 2.08                  & 9.70$\pm$0.81e-13                   & 30.40 \\
mean   & 8.15             &     & 0.36     & 1.60 & 1.66                  & 6.84e-13                            & 30.25 \\
\enddata
\tablenotetext{a}{
Notes: 
Mean count rate (Rate), counts, hardness ratio, median event energy (E$_{50}$), absorbed X-ray flux (F$_{x,abs}$), 
and unabsorbed X-ray luminosity (L$_{x}$, evaluated at d = 130.4 pc) are computed using events in the 
0.2-8 keV range. Hardness Ratio = counts(2-8 keV)/counts(0.2-8 keV). 
A (v) denotes significant count rate variability during the observation. 
Count rate uncertainties are 1$\sigma$. The values of the weighted plasma temperature
kT$_{wgtd}$, F$_{x,abs}$, and L$_{x}$  were determined
from 2T $apec$ thermal plasma models with fixed absorption N$_{\rm H}$ = 8e20 cm$^{-2}$
and metallicity $Z$ = 0.4 $Z_{\odot}$, except for ObsId 21166 where a 1T $apec$ model
gave an acceptable fit. For 2T $apec$ models, the value of kT$_{wgtd}$ is determined by
weighting the contribution of each temperature component by its respective XSPEC 
$norm$, or equivalenty by its volume emission measure (EM).
}
\tablenotetext{b}{Fit parameters are based on a 1T $apec$ model.}
\tablenotetext{c}{Fits of the high state spectrum (Fig. 2) using 143 events from the last 8.77 ks of 
the observation with a 2T $apec$ model give kT$_{cool}$ = 0.6($\pm$0.3) keV, 
kT$_{hot}$ = 2.4($\pm$0.7) keV, kT$_{wgtd}$ = 1.7 keV, 
F$_{x,abs}$ = 14.3($\pm$1.6)e-13 ergs cm$^{-2}$ s$^{-1}$, log L$_{x}$ = 30.58 ergs s$^{-1}$.}
\tablenotetext{d}{A Gaussian component at E = 1.473 keV was included in the spectral model used to fit the
full ObsId 21963 spectrum (low $+$ high states). This component improves the fit to a spectral feature
that is most likely the Mg XII doublet (E$_{lab}$ = 1.473 and 1.472 keV) which is visible in Figure 2.}
\end{deluxetable}

\normalsize

\clearpage

\section{Results}

\subsection{Undispersed X-ray Light Curves and Spectra}

V830  Tau was clearly detected in all four {\em Chandra} observations
as summarized in Table 3. Significant count rate variability was
detected only in ObsId 21963, for which the CIAO $glvary$ statistical 
test gives a probability of variability P$_{var}$ $>$ 0.999
based on arrival times of events in the 0.2-8 keV energy range.
The X-ray light curve (Fig. 1) shows an increase in count rate
occurring  $\approx$15 ks after the start of ObsId 21963. 
The mean count rate was 4.92 c ks$^{-1}$ 
during the first 14 ks (low-state) and  16.3 c ks$^{-1}$ during the 
last 8.8 ks (high-state) of the observation. The Hardness Ratio 
defined as H.R. = counts(2-8 keV)/counts(0.2-8 keV)
was H.R. = 0.22 during the first segment and H.R. = 0.43 during the 
second segment. A comparison of the low-state and high-state
ACIS-S 0-order spectra for low and high states is shown in Figure 2.
No significant variability was found in the other three observations but
the count rate and observed flux in the last observation (ObsId 21964) were 
about twice as high than in the first two observations. 

The source clearly became brighter during the 2.87 d interval 
over which the four observations were obtained. 
Since the 2.87 d interval spanned by all four observations 
covers only one stellar rotation period and 58\% of the planet's 
4.93 d orbital period reported by Donati et al. (2017), additional time 
monitoring over multiple cycles would be needed  to determine if the 
X-ray variability is tied to the rotational or orbital periods.  

\begin{figure}
\figurenum{1}
\includegraphics*[width=5.0cm,height=9.0cm,angle=-90]{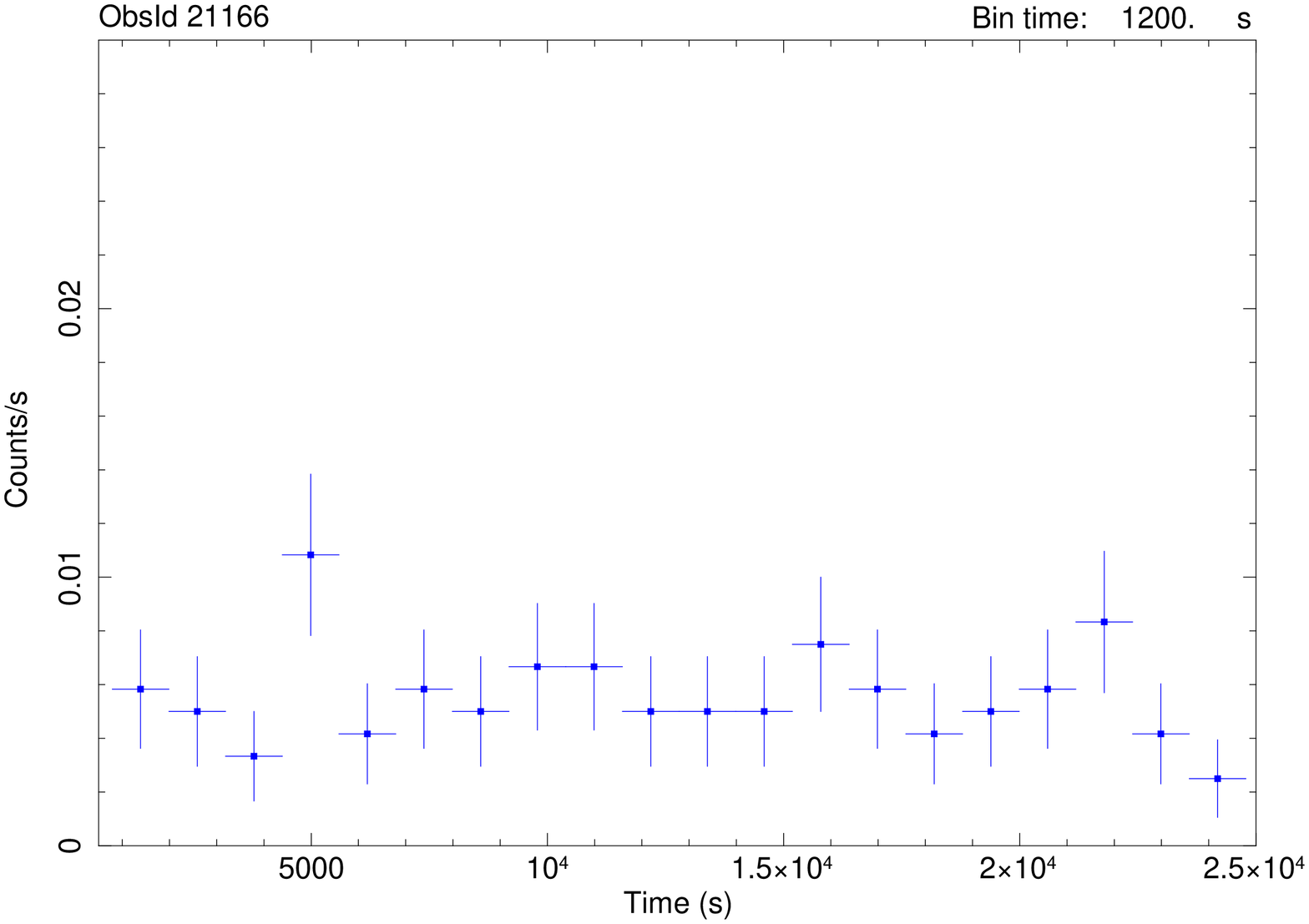} \\
\includegraphics*[width=5.0cm,height=9.0cm,angle=-90]{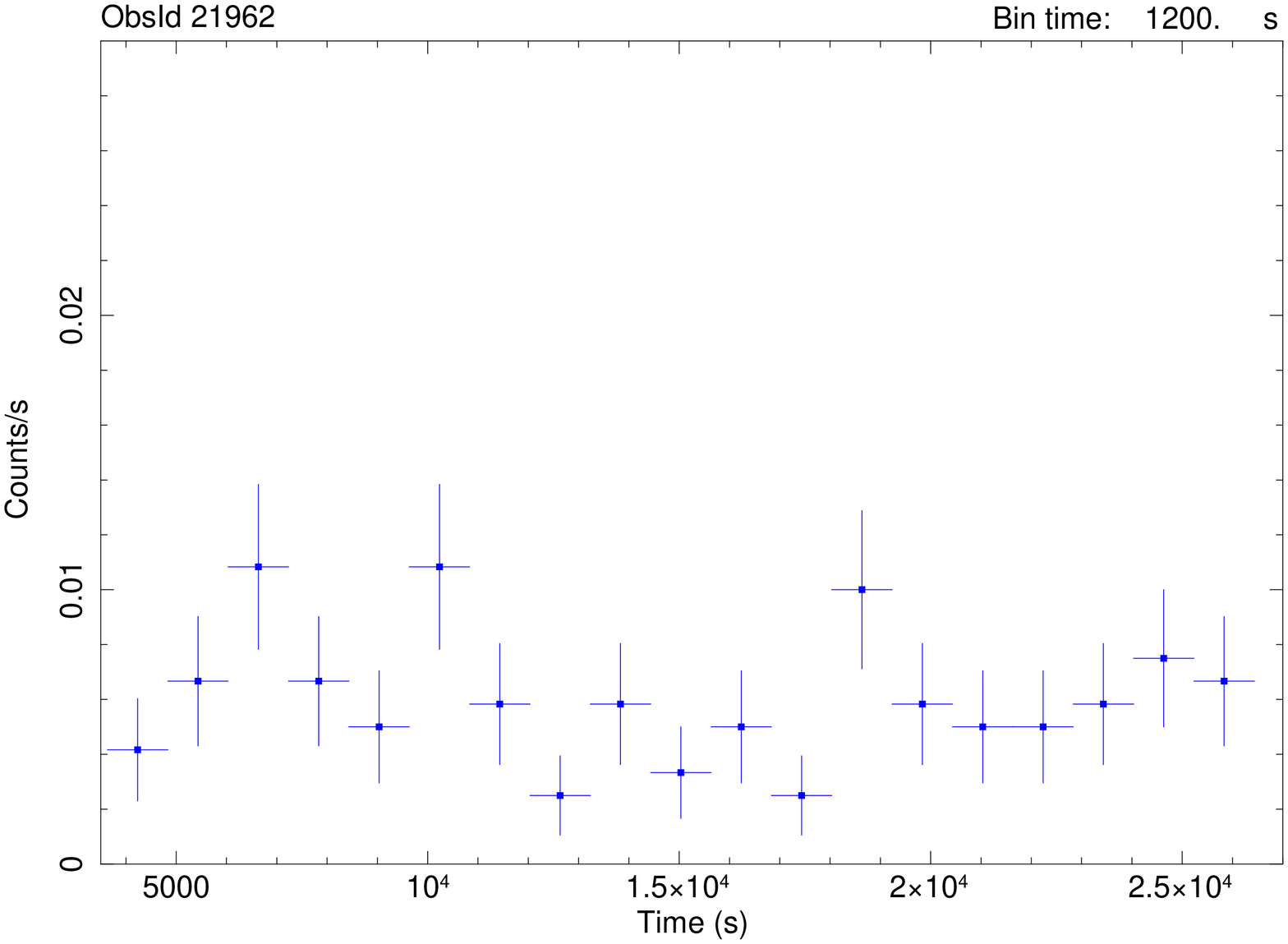} \\
\includegraphics*[width=5.0cm,height=9.0cm,angle=-90]{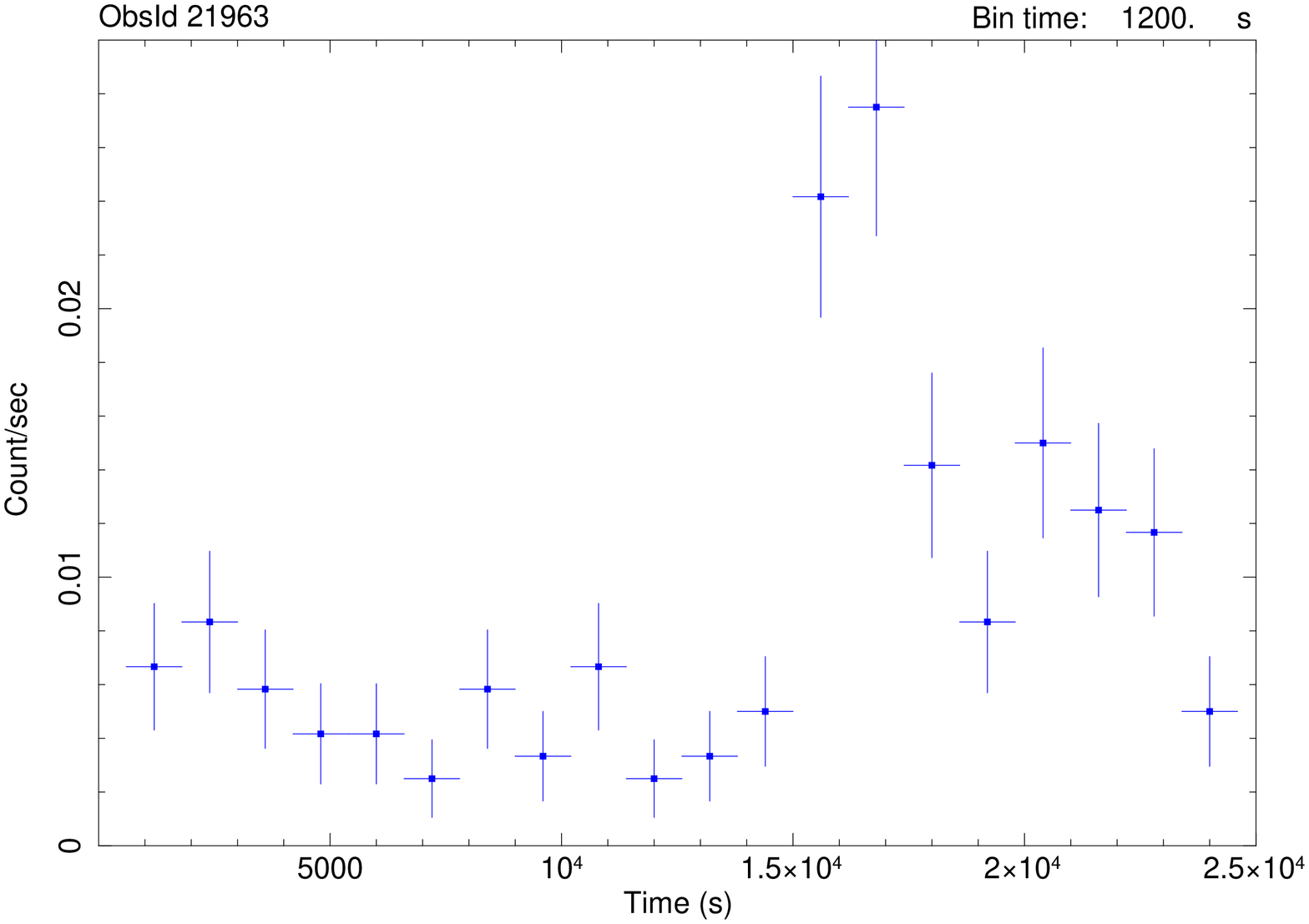} \\
\includegraphics*[width=5.0cm,height=9.0cm,angle=-90]{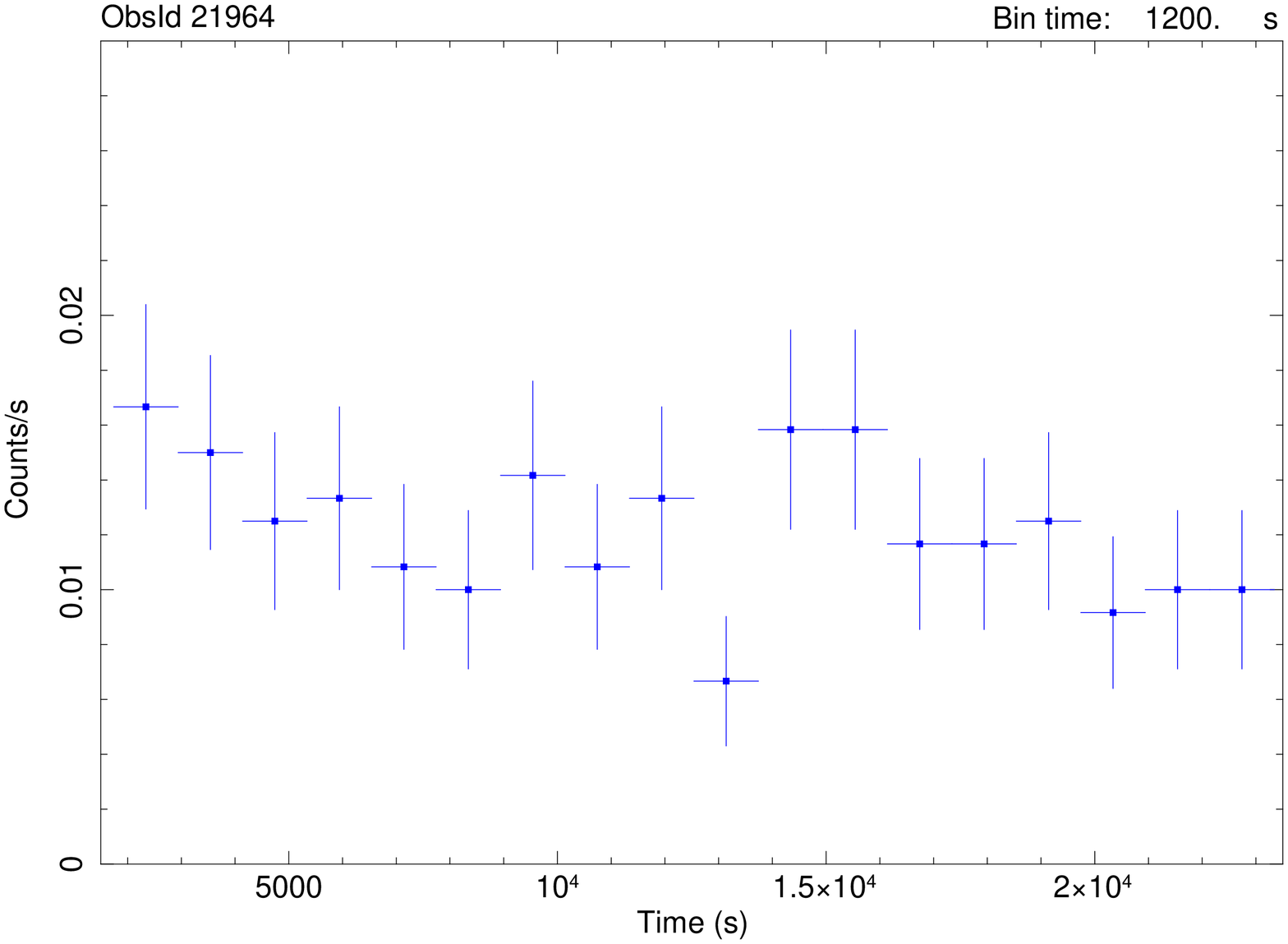} \\
\caption{{\em Chandra} ACIS-S 0-order light curves of V830 Tau  (0.2-8 keV)
         showing variability in ObsId 21963. All plots use the same Y-axis scale.
         Times are relative to start of each {\em Chandra} observation (Table 2).
}
\end{figure}

The combined 0-order spectrum consisting of events from all four 
observations is shown  in Figure 3-top. Since the source
varied over the course of the observations, the spectrum of each
observation was fitted separately.
The ACIS-S 0-order spectra were acceptably fitted with 
single-temperature (1T) or  two-temperature (2T)
thermal $apec$ plasma models (Smith et al. 2001), as summarized in Table 3.
The spectra do not have sufficient counts to reliably determine the 
absorption column density so it was held fixed at the value
N$_{\rm H}$ = 8 $\times$ 10$^{20}$ cm$^{-2}$ corresponding to
A$_{\rm V}$ = 0.4 mag and the conversion
N$_{\rm H}$ = 2 $\times$ 10$^{21}$A$_{\rm V}$ cm$^{-2}$
obtained by averaging the results of Gorenstein (1975)
and Vuong et al. (2003). 

The fits summarized in Table 3
use a metallicity fixed at $Z$ = 0.4 $Z_{\odot}$ since
the higher resolution grating spectra suggest some 
important metals such as Fe have subsolar abundances (Sec. 3.2).
The undispersed CCD spectra do not tightly constrain
the metallicity, which is sensitive to the temperature of
the hot plasma component (kT$_{2}$). A simultaneous fit of
the ACIS-S spectra for the first two observations where
no significant variability was detected allow values
$Z$ = (0.34 - 1.3) $Z_{\odot}$ (1$\sigma$ range), with smaller 
values of $Z$ corresponding to larger values of kT$_{2}$.
The assumed value of $Z$ has very little effect on the
value of L$_{x}$ determined from fits of the undispersed 
CCD spectra.

The first observation (ObsId 21166) has the 
softest spectrum and lowest observed flux, and it was
acceptably fitted using a 1T $apec$ model with a 
temperature kT = 1.10$\pm$0.08 keV.  During the remaining
three observations the  source flux and spectral hardness ratio 
increased.  These three observations were better fitted with
2T $apec$ models which gave kT$_{1}$ $\approx$ 0.7 keV and
kT$_{2}$ $\approx$ 2.2 - 3.8 keV. The last observation
(ObsId 21964) shows the hardest spectrum (H.R. = 0.47) and
its flux at energies above $\approx$2 keV is  
higher than the other three observations.

The results in Table 3 for ObsId 21963 when the source varied 
are based on all 212 events for the full observation.
In addition, separate spectra were extracted for the first 
14 ks during low-state  (69 counts) and last 8.8 ks during 
high-state (143 counts). In low-state the observed flux
was F$_{x,abs}$(0.2-8 keV) = 6.4($\pm$1.3)  $\times$ 10$^{-13}$
ergs cm$^{-2}$ s$^{-1}$ which increased to 
14.3($\pm$1.6)  $\times$ 10$^{-13}$ ergs cm$^{-2}$ s$^{-1}$ in high-state.

There are insufficient counts in the low-state spectrum to obtain
a reliable temperature measurement. Fits of the high-state spectrum
with 2T $apec$ models (Table 3 Notes) yield a hot component temperature
T$_{hot}$ = 28($\pm$8) MK). This value of T$_{hot}$ and the 
$\approx$9 - 10 ks decay time of the outburst are similar to those
reported  for the V830 Tau flare detected by {\em XMM-Newton}
(T$_{peak}$ $\geq$ 33 MK, $\tau_{decay}$ = 11.4 ks;
Franciosini et al. 2007).  By comparison, much more powerful flares
have been detected in T Tauri stars in the Taurus, Orion, and 
Rho Ophiuchi star-forming regions  with 
decay times $\tau_{decay}$ $\sim$ 100 ks and T$_{peak}$ $\sim$ 100 MK
(Franciosini et al. 2007; Getman et al. 2008; Imanishi et al. 2003).

The ACIS-S 0-order spectra show possible weak emission lines
or blended lines, most noticeable in the 0.9 - 1 keV range
from Ne IX and Ne X which are clearly visible in the 
MEG1 grating spectrum (Fig. 3). Line emission is discussed 
further in the grating analysis section below.

The mean intrinsic X-ray luminosity determined from ACIS-S
0-order fits at the {\em Gaia} EDR3 distance of 130.4 pc  is 
log L$_{x}$(0.2-8 keV)  = 30.25 ergs s$^{-1}$ but
varied over the range 30.10 - 30.40 ergs s$^{-1}$ (Table 3).
Fits of {\em Chandra} grating spectra allow slightly
higher L$_{x}$ values (Sec. 3.2).
The previous {\em XMM-Newton} observations measured a decline
in the X-ray luminosity of V830 Tau from
log L$_{x}$ = 30.87 to 30.39 ergs s$^{-1}$ with an
e-folding time of 11.4 ks, where we have adjusted the 
L$_{x}$ values reported by Franciosini et al. (2007) 
to d = 130.4 pc. The low end of the L$_{x}$ range  from  
{\em XMM-Newton} observations overlaps the
{\em Chandra} range and taken together the 
observations imply that L$_{x}$ can vary by at least 0.77 dex.

\begin{figure}
\figurenum{2}
\includegraphics*[width=9.0cm,angle=-90]{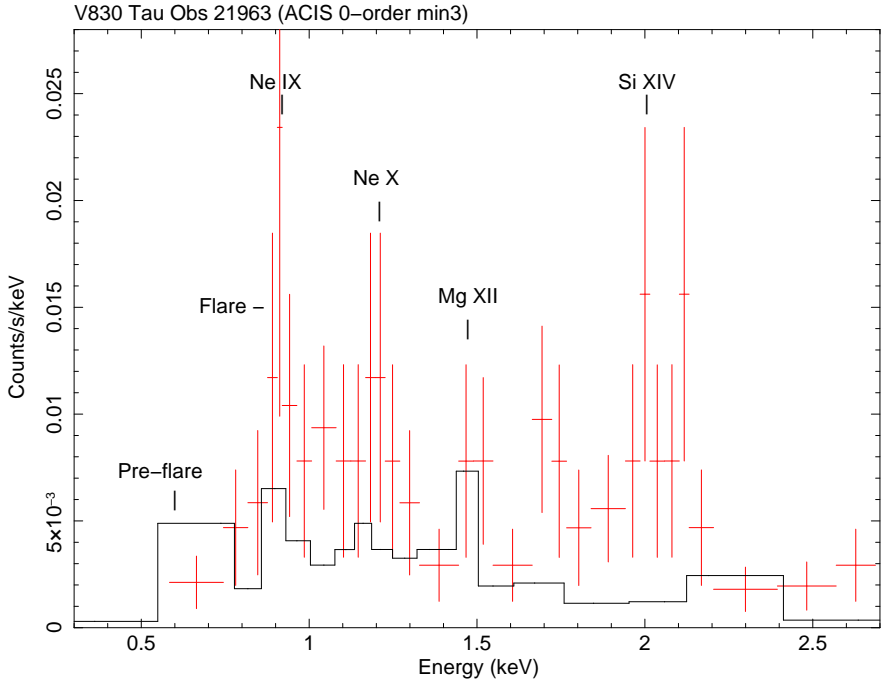} \\
\caption{
Comparison of the low-state (pre-flare, black) and high-state (flare, red) 
ACIS 0-order spectra of V830 Tau (ObsId 21963) lightly binned to a minimum of 
3 counts per bin. The low-state spectrum includes 69 counts from the first 14.2 ks 
of usable exposure and the high-state spectrum is based on 143 counts from the 
last 8.77 ks. Candidate emission lines are identified. The feature at 1.473 keV
is most likely weak Mg XII emission but it is not visible in the ObsId 21963 MEG1 spectrum.  
}
\end{figure}

\begin{figure}
\figurenum{3}
\includegraphics*[width=9.0cm,angle=-90]{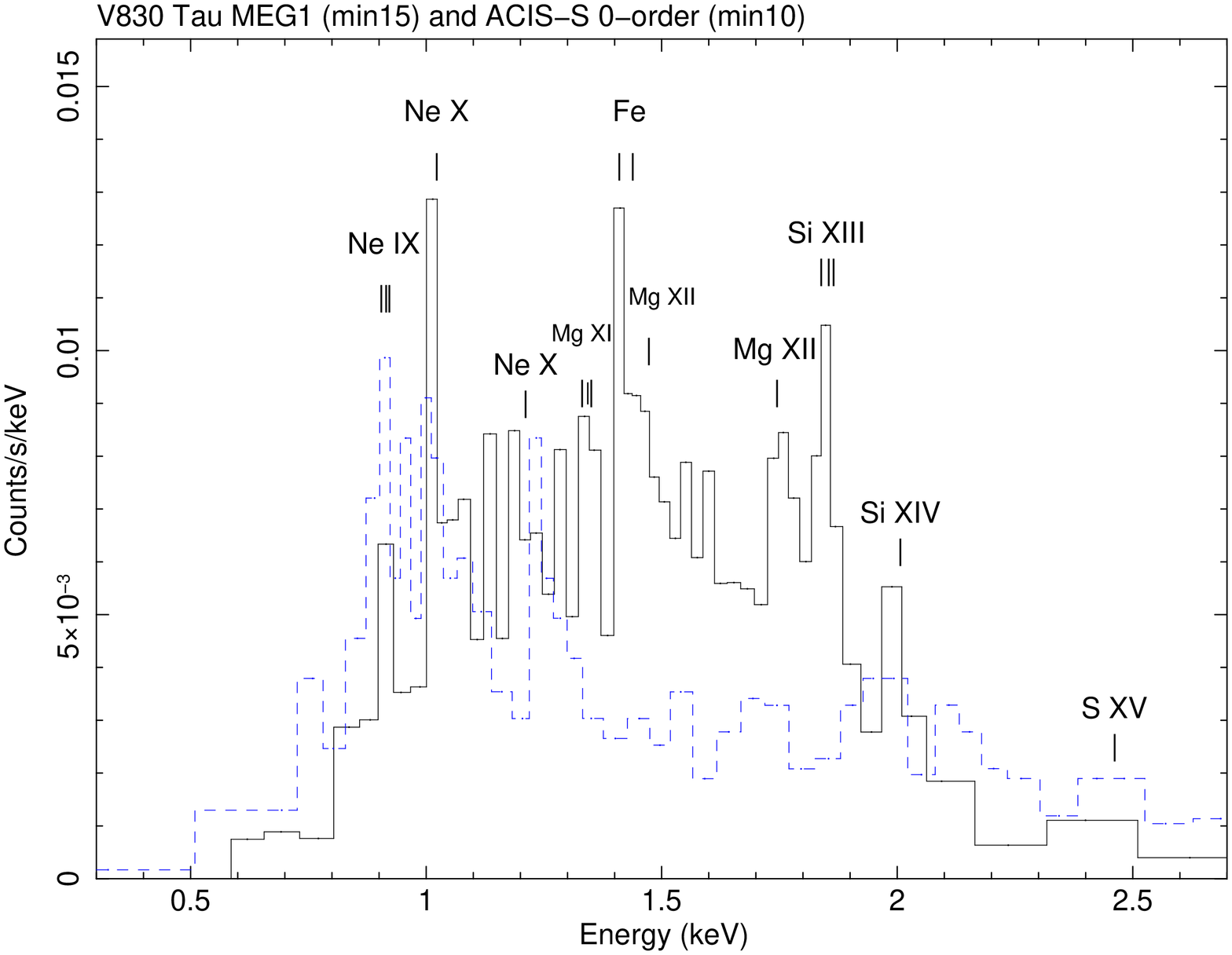} \\
\includegraphics*[width=9.0cm,angle=-90]{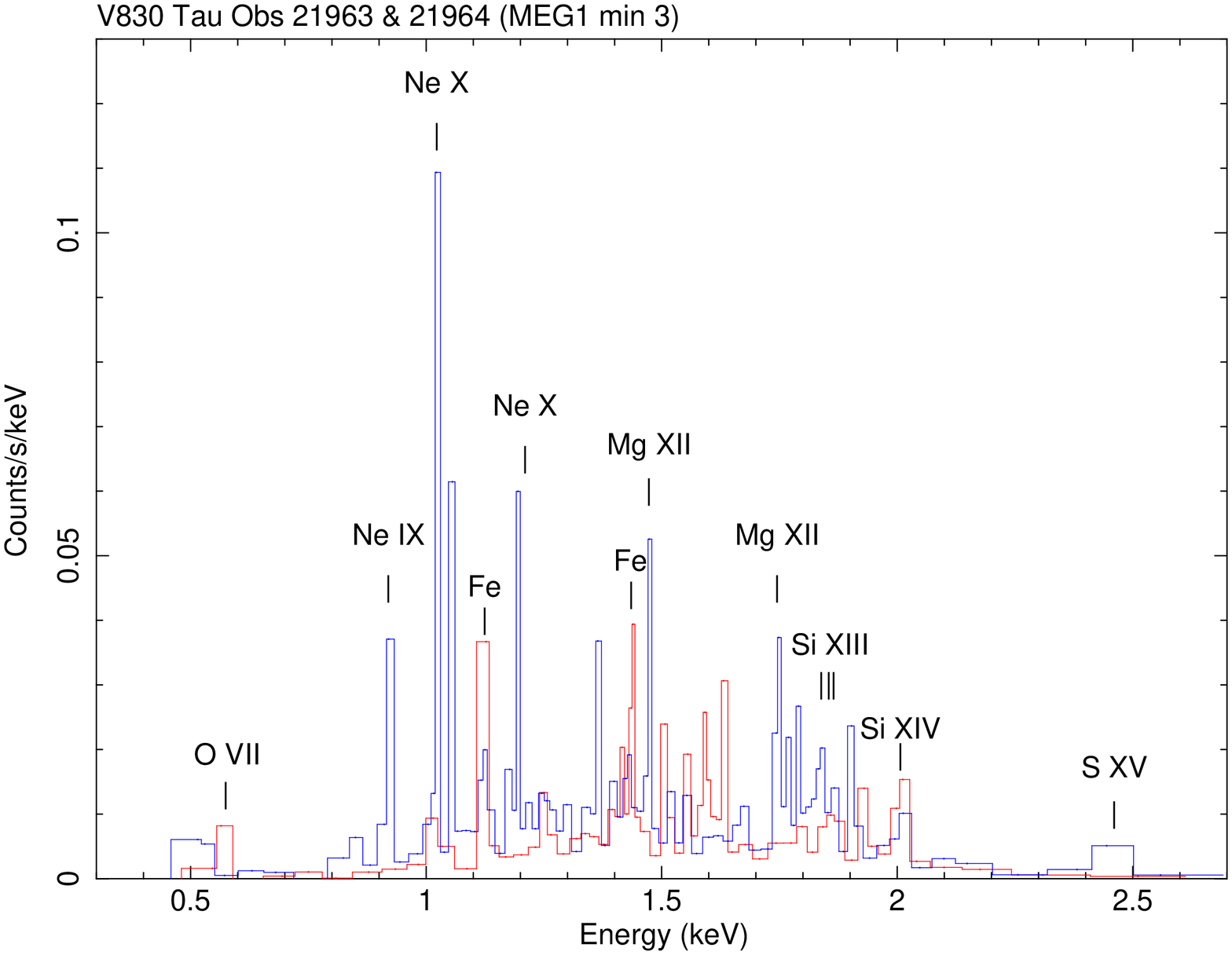} \\
\caption{
{\em Top}:~~Combined V830 Tau MEG1 spectrum of all four observations (solid line; 785 net counts)
binned to a minimum of 15 counts per bin. For comparison, the combined ACIS-S
0-order spectrum is shown (dashed line; 757 net counts; 10 counts per bin).
Error bars omitted for clarity.
{\em Bottom}:~
Overlay of V830 Tau MEG 1st order spectra ($+$1 and $-$1 orders combined)
for ObsIds 21963 (red) and 21964 (blue) on a linear energy scale. The spectrum has been
lightly binned in energy to a minimum of 3 counts per bin. Candidate emission
lines with lab energies (keV) are:  O VII$r$ (0.574), Ne IX$r$ (0.922), Ne X (1.022 and 1.211),
Fe XXIII/XXIV (1.124 - 1.129). Fe XXII/XXIII (1.423 - 1.439),
Mg XII (1.473 and 1.745), Si XIII$rfi$ (1.865, 1.854, 1.839), Si XIV (2.006),
S XV (2.461).
Error bars omitted for clarity.
}
\end{figure}
 

\begin{figure}
\figurenum{4}
\includegraphics*[width=9.0cm,angle=-90]{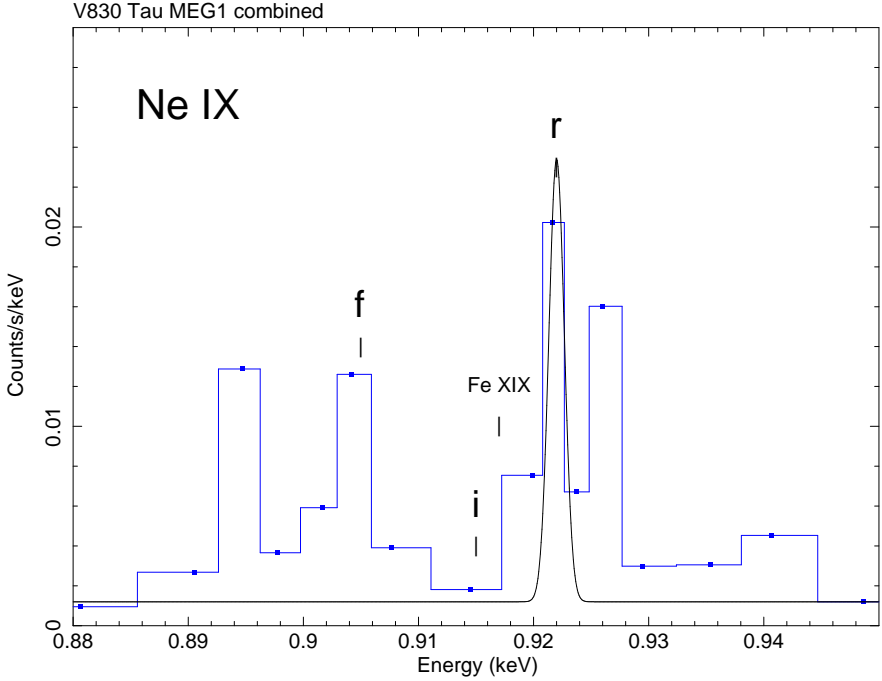} \\
\caption{V830 Tau MEG 1st order combined spectrum of the Ne IX
He-like triplet using data from  all four observations. The spectrum
has been lightly binned to a minimum of 2 counts per bin.
The vertical lines mark the lab energies of
the resonance ($r$), intercombination ($i$), and forbidden ($f$) lines at 
E$_{lab}$ = 0.922, 0.915 , and 0.905 keV. The lab energy of Fe XIX (0.917 keV) is also shown.
The Gaussian profile of the $r$ line uses a fixed centroid energy E = E$_{lab}$ = 0.922 keV and
width FWHM = 0.67 eV corresponding to 
MEG1 spectral resolution. The $i$ line is undetected. Error bars omitted for clarity.
}
\end{figure}

\subsection{Grating Spectra}

All four MEG1 spectra were  combined into a summed spectrum to bring out faint line 
emission, as shown in Figure 3-top. The brightest line detections
are summarized in Table 4, focusing mainly on  ObsIds 21963 and 21964 
(Fig. 3-bottom) which had higher count rates. Because of the low number
of counts in each line, line widths were held fixed at MEG1
instrumental resolution (FWHM $\Delta$$\lambda$ = 0.023 \AA~)
when fitting lines with Gaussian profiles. 

The lowest energy line is a possible detection of O VII 
at E$_{lab}$ = 0.574 keV (21.602 \AA~) which has maximum
line power at log T$_{max}$ = 6.3 (K) and is visible in
ObsId 21963 (Fig. 3-bottom). This feature is too faint for
a reliable flux measurement and is thus not included in Table 4.

The strongest lines are the Ne IX He-like triplet resonance line
at E$_{lab}$ = 0.922 keV (13.447 \AA~) shown in Figure 4 and the 
Ne X line at E$_{lab}$ = 1.022 keV (12.132 \AA~).
Neither line shows any significant centroid shift. Fitting the
MEG1 spectrum of ObsId 21964 using a fixed-width Gaussian
profile gives E$_{fit}$ = 0.919 ($\pm$0.001) keV for Ne IX$r$
and E$_{fit}$ = 1.021 ($\pm$0.002) keV for Ne X. 

In addition to the Ne IX$r$ line, the Ne IX$f$ 
forbidden line (0.905 keV) is  detected (ObsId 21964), but 
not the intercombination line Ne IX$i$ (0.915 keV). ObsId 21964
shows the highest Ne IX flux and a 3-component Gaussian fit
gives a relative flux normalization of the forbidden to resonance
lines $r$/$f$ = 1.65/1.10 = 1.5. The $i$ line flux normalization is
consistent with zero. Thus, no indication of a suppressed 
flux ratio $R$ = $f$/$i$ that can result from high electron densities
is present. No significant flux was detected for the Mg XI or 
Si XIII triplets  in the individual observations but a measurement 
of the summed $r+i+f$ fluxes in the combined MEG1 spectrum (Table 4) 
indicates that Mg XI emission is present as well as weaker 
emission from Si XIII. 
The highest temperature lines detected are Si XIV (E$_{lab}$ = 2.006 keV) 
and S XV (E$_{lab}$ = 2.461 keV), both of which have maximum line power
at  T$_{max}$ = 15.8 MK. There is also weak emission at S XV in the
combined ACIS 0-order spectrum (Fig. 3-top). 
 
There are insufficient counts in the MEG1 spectra of the 
individual observations to obtain useful abundance constraints.
Thus the combined spectrum (785 net counts) using data from all four observations 
was fitted using a 2T $vapec$ variable abundance model.
The abundances of elements with detected emission lines (Ne, Mg, Si, S, Fe) 
were allowed to vary either individually or in combination, but the abundance of each 
element was constrained to be equal for the cool and hot plasma components. 
The abundances of other elements in the $vapec$ model (C, N, O, Al, Ar, Ca, Ni) 
were all held fixed at the same value, which was taken to be either solar  
or subsolar (values of 0.2 or 0.4 $\times$ solar were compared).
The column density was held fixed at N$_{\rm H}$ = 8 $\times$ 10$^{20}$ cm$^{-2}$.  

The results of several different variable abundance fits can be summarized as
follows: (i) abundances of varied elements converged to subsolar
values (most notably Fe), as was also generally found for TTS in the Taurus 
molecular cloud  observed in the {\em XEST} study (G\"{u}del et al. 2007),
(ii) fit results were insensitive to the S abundance, 
(iii) cool and hot plasma component temperatures ranged
from kT$_{cool}$ = 0.37 - 0.43 keV and kT$_{hot}$ = 2.0 - 2.6 keV,
(iv) the absorbed flux was in the range F$_{x,abs}$(0.2-8 keV) = 
(1.1 - 1.2) $\times$ 10$^{-12}$ ergs cm$^{-2}$ s$^{-1}$, and 
(v)  luminosity log L$_{x}$ = 30.51 - 30.59 ergs s$^{-1}$.  This L$_{x}$
range is slightly above that obtained from fits of the ACIS-S CCD
spetra and may reflect better accounting of line fluxes in the
MEG1 spectrum. A representative 2T $vapec$ variable abundance fit is 
summarized in Table 5.


\begin{deluxetable}{lcccc}
\tablewidth{0pt}
\tablecaption{V830 Tau X-ray Line Fluxes (MEG1)}
\tablehead{
           \colhead{Ion}            &
           \colhead{E$_{lab}$}      &
           \colhead{log T$_{max}$}  &
           \colhead{ObsId}          &
           \colhead{Net Line Flux}  \\
           \colhead{}               &
           \colhead{(keV)}           &
           \colhead{(K)}            &
           \colhead{}               &
           \colhead{(10$^{-6}$ ph cm$^{-2}$ s$^{-1}$)}             
}
\startdata
 Ne IX  & 0.922  & 6.6  & 21964  & ~~~~28.4 (15.3)\tablenotemark{a,b} \\
 Ne X   & 1.022  & 6.8  & 21166  & 15.4 (7.5) \\
 Ne X   & 1.022  & 6.8  & 21964  & 15.1 (7.7) \\
 Ne X   & 1.211  & 6.8  & 21166  & 3.3  (2.1) \\
 Ne X   & 1.211  & 6.8  & 21962  & 2.5  (1.8) \\
 Mg XI  & 1.352  & 6.8  & sum\tablenotemark{c} & ~4.4 (1.1)\tablenotemark{a} \\
 Mg XII & 1.473  & 7.0  & 21964  & 3.3 (1.4) \\
 Mg XII & 1.745  & 7.0  & 21964  & 3.2 (1.2) \\
Si XIII & 1.865  & 7.0  & sum\tablenotemark{c} & ~2.6 (0.6)\tablenotemark{a} \\
Si XIV  & 2.006  & 7.2  & 21963  & 3.0 (1.3) \\
Si XIV  & 2.006  & 7.2  & 21964  & 2.1 (1.1) \\
 S XV   & 2.461  & 7.2  & 21964  & 5.8 (2.8) 
\enddata
\noindent Notes: 
Laboratory energies (E$_{lab}$) and maximum line power temperatures (T$_{max}$) are from
the ATOMDB database (www.atomdb.org). For He-like triplets (Ne IX, Mg XI, Si XIII)
E$_{lab}$ is for the resonance ($r$) line.
Net photon (ph) line fluxes are continuum-subtracted with uncertainties in parentheses. 
Gaussian line widths were held fixed at MEG1 spectral resolution (0.023 \AA) during fits.
\tablenotetext{a}{Sum of the triplet components:
resonance ($r$) $+$ intercombination ($i$) $+$ forbidden ($f$) lines
based on a 3-component Gaussian fit.}
\tablenotetext{b}{Triplet line fluxes are
$r$ = 16.6, $i$ $\leq$ 1.5, $f$ = 11.8.}
\tablenotetext{c}{Only the summed spectrum of all four observations 
                  procided a measurable line flux. Thus, the quoted
                   value is an average over the summed exposure of 90.245 ks (Table 2).}
\end{deluxetable}
\clearpage


\begin{deluxetable}{cccccccc}
\tabletypesize{\footnotesize}
\tablewidth{0pt}
\tablecaption{V830 Tau MEG1 Spectral Fit}
\tablehead{
           \colhead{N$_{\rm H}$}          &
           \colhead{kT$_{1}$}             &
           \colhead{kT$_{2}$}             &
           \colhead{norm$_{1}$}           &
           \colhead{norm$_{2}$}           &
           \colhead{Abundances}           &
           \colhead{F$_{x}$}              &
           \colhead{log L$_{x}$}       \\
           \colhead{(cm$^{-2}$)}            &
           \colhead{(keV)}                  &
           \colhead{(keV)}                  &
           \colhead{(cm$^{-5}$)}            &
           \colhead{(cm$^{-5}$)}            &
           \colhead{}                       &
           \colhead{(ergs cm$^{-2}$ s$^{-1}$)}  &
           \colhead{(ergs s$^{-1}$)}        
}
\startdata
 8e20 & 0.42 [0.28 - 1.23] & 2.07 [1.12 - 6.34]  & 14.0e-04  & 5.9e-04 & varied\tablenotemark{a} & 1.11 (1.62)e-12  & 30.52    \\
\enddata
\tablecomments{
Based on 2T $vapec$ fits of the combined MEG1 spectrum (785 net counts) from all four 
observations binned to a minimum of 10 counts per bin. The column density 
N$_{\rm H}$ was held fixed, The abundances of Ne, Mg, Si, Fe were allowed to vary.
The abundances of other metals were held fixed at an abundance 0.4 $\times$ solar.
Abundances are relative to solar photospheric values of Anders \& Grevesse (1989).
Brackets enclose 1$\sigma$ confidence ranges.
The XSPEC $norm$ is related to the volume emission measure by
EM = $n_{e}^2$V = 4$\pi$ $\times$ 10$^{14}$d$_{cm}^2$$\cdot$norm where 
$n_{e}$ is the electron density in the X-ray emitting plasma of volume V
and d$_{cm}$ is the source distance in cm. For V830 Tau, d = 130.4 pc
gives $n_{e}^2$V = 2.03 $\times$ 10$^{42}$$\cdot$norm. 
The flux F$_{x}$(0.2 - 8 keV) is the absorbed (observed) value followed in parentheses
by the unabsorbed value.
}
\tablenotetext{a}{The varied abundances converged to
Ne =  0.25 [0.04 - 0.88], Mg = 0.15 [* - 0.48], Si = 0.23 [0.06 - 0.58], Fe = 0.07 [* - 0.33] $\times$ solar.
An asterisk denotes no convergence on the lower bound by the error calculation algorithm.
}
\end{deluxetable}


\section{Discussion}

\subsection{Comparison of V830 Tau with Other T Tauri Stars in Taurus}

Given the possible presence of a close-in planet orbiting V830 Tau
it is appropriate to ask whether its X-ray luminosity is anomalous
compared to other weak-lined TTS. Using data acquired during the
{\em XMM-Newton} Extended Survey of the Taurus Molecular Cloud (XEST),
Telleschi et al. (2007) found a high probability of correlation
between L$_{x}$ and  stellar luminosity L$_{*}$ in T Tauri stars.
Using a Kaplan-Meier
estimator that takes both detections and upper limits into account
they obtained a mean value for weak-lined TTS in the XEST sample
log(L$_{x}$/L$_{*}$) = $-$3.36$\pm$0.07. 

To compare this with
the {\em Chandra} observations of V830 Tau we use 
log L$_{x}$ = 30.42$\pm$0.17 ergs s$^{-1}$ based on an average of the 
mean value  obtained from the ACIS-S 0-order fits (Table 3) and the 
somewhat higher value obtained from the combined MEG1 fit  
(Table 5). Taking L$_{*}$ = 1.2$\pm$0.3 L$_{\odot}$ (Table 1) for V830 Tau gives
log(L$_{x}$/L$_{*}$) = $-$3.24 ($-$3.51 - $-$2.96) where the range
in parentheses accounts for the spread in L$_{x}$ and the 
uncertainty of L$_{*}$. This value is slightly higher than the 
XEST mean by 1.7$\sigma$ but the range in parentheses overlaps
the XEST mean value. Thus, the {\em Chandra} observations of
V830 Tau do not deviate significantly from the L$_{x}$/L$_{*}$
ratio obtained from the XEST sample.  On the other hand, the higher value 
log L$_{x}$ = 30.87 ergs s$^{-1}$ from the previous {\em XMM-Newton}
observation gives log(L$_{x}$/L$_{*}$) = $-$2.79 which is
8.6$\sigma$ higher than the XEST mean. So during periods of
strong activity (e.g. flaring) V830 Tau's X-ray emission is
well above the normal range for weak-lined TTS in Taurus.

\subsection{X-ray Ionization and Heating}

X-ray and EUV radiation ionizes and heats the planet's atmosphere.
The X-ray ionization rate $\zeta$ at a given X-ray optical depth $\tau_{x}$ 
in the planet's atmosphere depends on the star's X-ray luminosity, 
plasma temperature kT$_{x}$, and the separation $a$.
The X-ray heating rate $\Gamma_{x}$ depends on $\zeta$
and the planet's atmospheric properties,  
i.e the number density n$_{\rm H}$ at the height where
the heating rate is computed. Details on  calculation
of the ionization and heating rates are given below.

Absorption of an X-ray photon of energy E by gas in the planet's 
atmosphere results in ionization and production of a primary 
photoelectron having energy E - E$_{i}$ where  E$_{i}$ is
the ionization potential. The primary photoelectron in turn
produces multiple secondary ionizations which dominate the 
X-ray heating.
The X-ray photoelectric absorption cross-section drops off rapidly with
incident photon energy and is approximated by a power-law
$\sigma(E)$ = $\sigma_{0}$(E/1 keV)$^{-p}$ cm$^{-2}$ where we use
$\sigma_{0}$ = 2.27 $\times$ 10$^{-22}$ cm$^{2}$ and $p$ = 2.485
as appropriate for solar abundance gas (Igea \& Glassgold 1999;
Shang et al. 2002). If metals are depleted onto grains then
the value of $p$ must be adjusted (Glassgold et al. 1997).

The X-ray optical depth for a photon of energy E is related to the 
equivalent neutral H column density via $\tau_{x}$ = $\sigma(E)$N$_{\rm H}$.
At unit optical depth N$_{\rm H}$($\tau_{x}$=1) =  1/$\sigma(E)$.
For the median photon energy E$_{50}$ = 1.6 keV of
V830 Tau the above relation gives  $\tau_{x}$ = 1 at
N$_{\rm H}$ = 1.4 $\times$ 10$^{22}$ cm$^{-2}$.
If $n_{\rm H}$($z$) is the number density at height $z$ above the 
planet's surface then the corresponding column density is
N$_{\rm H}$ = $\int$$n_{\rm H}(z)$d$z$ where the integral is evaluated
along the line-of-sight from the star to the target point in the atmosphere.
An atmospheric model specifying $n_{\rm H}$($z$) is needed to determine
the  height $z$ corresponding to a given value of $\tau_{x}$ (Sec. 4.4).

At X-ray energies E $\ltsimeq$ 3 keV that are relevant for V830 Tau,
scattering effects are negligible compared to photoelectric
absorption and the scattering cross-section can be ignored (Bruderer et al. 2009).
Because of the steep decline of $\sigma(E)$ with energy,
low-energy X-ray photons (and EUV photons) will be absorbed
higher in the planet's atmosphere and the absorbed spectrum
thus becomes harder and weaker with increasing penetration depth
as illustrated in more detail below 
(see also Cecchi-Pestellini et al. 2006).

We use the analytical development of Shang et al. (2002) to compute
the X-ray ionization and heating rates. At a distance $r$ from the 
star, the total X-ray ionization rate corresponding to a primary 
ionization rate $\zeta_{\rm x}$ in a thermal
plasma at X-ray temperature kT$_{x}$ is

\begin{equation}
\zeta \approx \zeta_{\rm x} \left[{ \frac{r}{R_{\rm x}}} \right]^{-2} \left[{ \frac{kT_{x}}{\epsilon_{ion}}} \right] I_{p}(\tau_{\rm x})~~~({\rm s}^{-1}~{\rm per~ H~nucleus)}.
\end{equation}
In the above $R_{\rm x}$ $\geq$ R$_{*}$ sets the distance of the X-ray source 
above the star, $\epsilon_{ion}$ $\approx$ 37 eV  is the energy to create an 
ion pair (i.e. the net energy of the primary photoionized electron E $-$ E$_{i}$
divided by the number of secondary ionizations it produces), 
and the function $I_{p}(\tau_{\rm x})$ gives
the X-ray attenuation at optical depth $\tau_{\rm x}(r,E)$,
as described in Appendix C of Shang et al. (2002).
For X-rays originating in coronal loops, one expects $R_{\rm x}$ to be a few 
stellar radii but for the calculations given below we simply set
$R_{\rm x}$ = R$_{*}$ = 2 R$_{\odot}$. Assuming a planet at separation
$r$ = $a$ = 0.057 au (Donati et al. 2017) gives  
$r$/R$_{x}$ = $r$/R$_{*}$ = 6.14.

The primary X-ray ionization  rate is (Shang et al. 2002)

\begin{equation}
\zeta_{\rm x} =  \frac{L_{x}\sigma(kT_{x})}{4 \pi R_{x}^2 kT_{x}}  =  1.13 \times 10^{-8}  \left[{ \frac{L_{x}}{10^{30}~ {\rm erg~ s}^{-1}}} \right] \left[{ \frac{kT_{\rm x}}{{\rm keV}}} \right]^{-(p+1)} \left[{ \frac{R_{\rm x}}{10^{12}~{\rm cm}}} \right]^{-2}~~~({\rm s}^{-1})
\end{equation}
where, as above, $\sigma$(kT$_{x}$) =  $\sigma(E)$  is the photoelectric X-ray absorption cross-section
per H nucleus evaluated at energy $E$. We compute
$\zeta$ as a function of $\tau_{x}$ at 
$r$ = 0.057 au, adopting a plasma temperature kT$_{x}$ = 1.6 keV,
and log L$_{x}$ = 30.4 ergs s$^{-1}$. The adopted values of kT$_{x}$ and L$_{x}$ 
are  representative of the range determined from existing 
{\em Chandra} and {\em XMM-Newton} observations of V830 Tau (Table 4; Sec. 3.1). 
The computed ionization rate as a function of $\tau_{x}$ is shown in Figure 5
and values corresponding to $\tau_{x}$ = 1 at 1.6 keV are given in Table 6.
Figure 6 shows a comparison of the incident X-ray spectrum on the planet's
atmosphere and the absorbed spectrum at
N$_{\rm H}$ = 1.4 $\times$ 10$^{22}$ cm$^{-2}$ ($\tau_{x}$ = 1 at 1.6 keV).

The  X-ray heating rate per unit volume is proportional to the ionization rate
and is given by 

\begin{equation}
\Gamma_{\rm x} = \zeta n_{\rm H} Q
\end{equation}
where $Q$ is the heating rate per ionization and
$n_{\rm H}$ is the number density of hydrogen nuclei in the planet's
atmosphere at the height corresponding to the optical depth 
$\tau_{x}$ at which $\zeta$ was computed (eq. 1). The value of Q depends on
several factors including the nature of the gas (i.e. atomic versus
molecular), as discussed by Glassgold et al. (2012). Using a 
fiducial value Q = 20 eV gives 

\begin{equation}
\Gamma_{\rm x} = 3.2 \times 10^{-11} \zeta  n_{\rm{H}}\left[\frac{Q}{20~{\rm eV}}\right]~~~{\rm (ergs~cm^{-3}~s^{-1}) }            .
\end{equation}

The representative value of $\Gamma_{\rm x}$ in Table 6 leaves $n_{\rm H}$ as a free
parameter since it depends on the planet's atmospheric structure, about
which nothing is yet known.

\begin{deluxetable}{cccccc}
\tabletypesize{\footnotesize}
\tablewidth{0pt}
\tablecaption{X-ray Ionization and Heating Rates (V830 Tau)}
\tablehead{
           \colhead{r}               &
           \colhead{kT$_{x}$}        &
           \colhead{$\zeta_{x}$}     &
           \colhead{$\zeta$}         &
           \colhead{$\tau_{x}$}      &
           \colhead{$\Gamma_{x}$}      \\
           \colhead{(au)}            &
           \colhead{(keV)}           &
           \colhead{(s$^{-1}$)}      &
           \colhead{(s$^{-1}$ H$^{-1}$)}   &
           \colhead{}                &
           \colhead{(ergs s$^{-1}$ cm$^{-3}$ $n_{\rm H}^{-1}$)} 
 }
\startdata
  0.057 & 1.6 & 2.83e-07 & 3.43e-08 & 1   & 1.10e-18  \\
\enddata
\tablecomments{
The  secondary ionization rate $\zeta$ (eq. [1]), 
primary ionization rate $\zeta_{x}$ (eq. [2]), and
heating rate $\Gamma_{x}$ (eq. [4]) are computed 
at $\tau_{x}$ = 1 using
kT = 1.6 keV, L$_{x}$ = 2.5 $\times$ 10$^{30}$ ergs s$^{-1}$,
R$_{x}$ = R$_{*}$, Q = 20 eV, and an assumed planet separation $r$ = $a$ = 0.057 au.
The value of $n_{\rm H}$ required to evaluate $\Gamma_{x}$ depends on the
adopted planet atmosphere model and has been left as a free parameter.
}
\end{deluxetable}
\clearpage

\begin{figure}
\figurenum{5}
\includegraphics*[width=9.0cm,angle=-90]{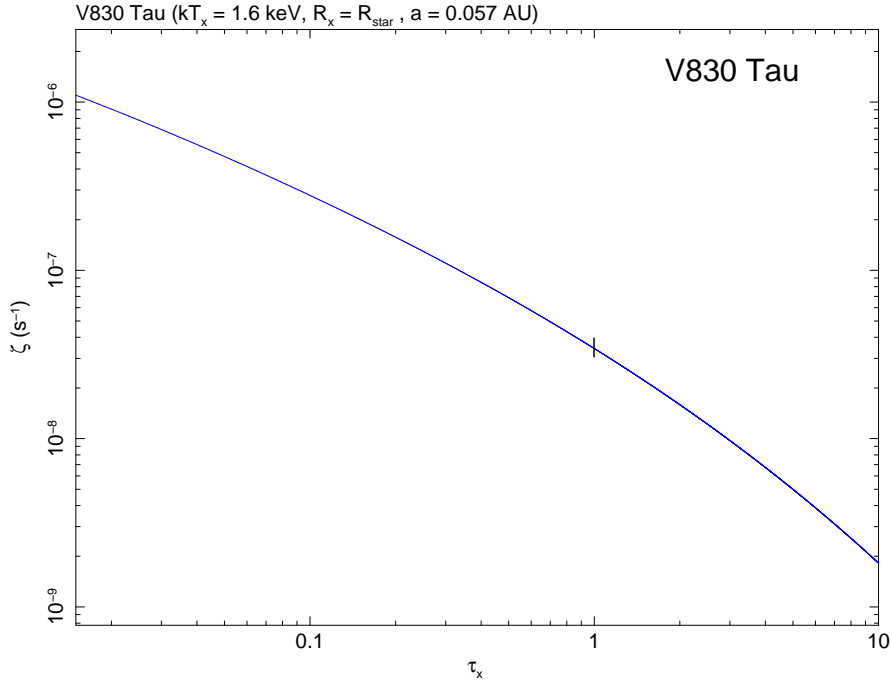} \\
\caption{X-ray ionization rate versus X-ray optical depth computed at 
at a distance 0.057 au from V830 Tau for a stellar
X-ray temperature kT$_{x}$ = 1.6 keV and luminosity
log L$_{x}$ = 30.4 ergs s$^{-1}$ (Eq. 1; Table 6).  
}
\end{figure}


\begin{figure}
\figurenum{6}
\includegraphics*[width=9.0cm,angle=-90]{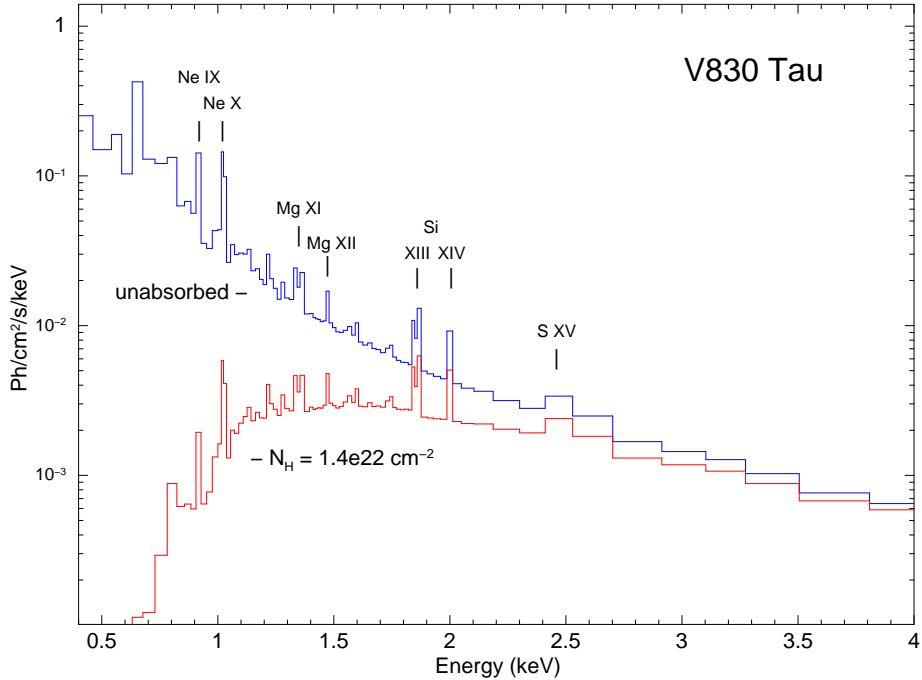} \\
\caption{Unfolded {\em Chandra} MEG1 spectrum of V830 Tau based on a
2T $vapec$ variable abundance model (Table 5). The unabsorbed 
(N$_{\rm H}$ = 0) version of the model depicts the shape of the spectrum
incident on the planet. The absorbed model (N$_{\rm H}$ = 1.4e22 cm$^{-2}$)
corresponds to $\tau_{x}$ = 1 at E = 1.6 keV and shows low-energy
photon absorption by the planet's atmosphere. The photon flux density
has been normalized to unity for display.
}
\end{figure}
\clearpage

\subsection{EUV Ionization and Heating}

EUV photons ($\lambda$ = 124 - 920 \AA) have  energies  
E = 0.013 - 0.1 keV, less energetic than X-rays but still
capable of ionizing hydrogen and heating the planet's outer atmosphere.
Because of their lower energies, EUV photons  are absorbed  higher in the
atmosphere than X-ray photons. At column densities  
N$_{\rm H}$ $\gtsimeq$ 10$^{19}$ - 10$^{20}$ cm$^{-2}$
the optical depth at EUV energies becomes large and
EUV heating is negligible (Cecchi-Pestellini et al. 2009).
For  T Tauri stars with strong EUV emission there is 
a balance  between EUV photoionization and radiative recombination
at optical depths $\tau_{EUV}$ $<$ 1, and Ly$\alpha$ cooling in the
planet's atmosphere must be taken into account.

The EUV heating rate $\Gamma_{EUV}$ is computed in a manner similar 
to that of X-rays (Cecchi-Pestellini et al. 2009; Murray-Clay et al. 2009). 
For a simplified monoenergetic EUV flux at energy E$_{0}$ the heating
rate is

\begin{equation}
\Gamma_{\rm EUV} = \epsilon {\rm F}_{EUV} \sigma(E_{0}) {\rm e}^{-\tau_{\rm euv}}  n_{\rm{H}}~~~{\rm (ergs~cm^{-3}~s^{-1}) }  
\end{equation}

\noindent where $\epsilon$ is the fraction of primary photoelectron energy that 
goes into gas heating, F$_{EUV}$ is the 
unattenuated EUV energy flux impinging on the planet's atmosphere, 
$\sigma(E_{0}$) is the EUV photoelectric cross-section, $\tau_{\rm euv}$ = $\sigma(E_{0}$)N$_{\rm H}$,
and n$_{\rm H}$ is the number density corresponding to N$_{\rm H}$ at 
optical depth $\tau_{\rm euv}$.  The fractional heating efficiency
$\epsilon$ $<$ 1 depends on the electron density and values as small 
as $\epsilon$ $\approx$ 0.1 - 0.2 are possible (Cecchi-Pestellini et al. 2009).
The ionization rate 
$\zeta_{EUV}$ = F$_{EUV}$e$^{-\tau_{euv}}$ $\sigma$(E$_{0}$)/E$_{0}$~(s$^{-1}$ per H nucleus)
is incorporated into Eq. (5). 

At a distance $a$ from the star F$_{EUV}$ = L$_{EUV}$/(4$\pi$$a^2$).
Using L$_{x}$ = 2.5 $\times$ 10$^{30}$ ergs s$^{-1}$
for V830 Tau, Eq. (5) evaluated at $a$ = 0.057 au can be written as

\begin{equation}
\Gamma_{\rm EUV} = 2.75 \times 10^{5} \epsilon \left[\frac{L_{EUV}}{L_{x}}\right] \sigma(E_{0}) {\rm e}^{-\tau_{\rm euv}}  n_{\rm{H}}~~~{\rm (ergs~cm^{-3}~s^{-1}) }
\end{equation}

To numerically compute $\Gamma_{EUV}$, the flux F$_{EUV}$ 
(or equivalently L$_{EUV}$) must be known.
However, stellar EUV fluxes cannot be measured except for 
the Sun and a few nearby stars because of strong absorption by neutral 
hydrogen in the interstellar medium. Analysis of samples of late-type stars of
various ages yield estimates of L$_{EUV}$/L$_{x}$, but with 
large scatter. For the young solar-type star EK Dra, 
Ribas et al. (2005) obtained L$_{EUV}$ $\approx$ L$_{x}$,
whereas other studies based on a larger sample of stars 
give L$_{EUV}$ $\sim$ a few times L$_{x}$ (e.g. Sanz-Forcada et al. 2011).
Methods for estimating L$_{EUV}$ have also been obtained using
correlations between L$_{x}$ and L$_{Ly\alpha}$
(Linsky et al. 2013; 2015). 
In the absence of an observational measurement some general 
studies of planetary atmospheres assume L$_{EUV}$ $\sim$ L$_{x}$ 
(e.g. Owen \& Jackson 2012). Since  L$_{x}$ is variable for
V830 Tau we also expect L$_{EUV}$ to vary, as is true even
for older relatively inactive stars like the Sun (Woods et al. 2005). 

For the EUV cross section we adopt $\sigma$(E = 0.1 keV) = 6 $\times$ 10$^{-20}$ cm$^{2}$
(Bruderer et al. 2009) as a reference value and a profile
$\sigma$(E) = 6 $\times$ 10$^{-20}$(E/100 eV)$^{-p}$ (cm$^{2}$) with $p$ = 2.485. 
At a representative EUV energy  E$_{0}$ = 60 eV the cross-section is
$\sigma$(60 eV) $\approx$ 2 $\times$ 10$^{-19}$ cm$^{2}$
and $\tau_{\rm EUV}$ = 1 corresponds to N$_{\rm H}$ = 5 $\times$ 10$^{18}$ cm$^{-2}$.   
At this energy the heating rate at unit optical depth becomes
$\Gamma_{EUV}$ = 2 $\times$ 10$^{-14}$$\epsilon$[L$_{EUV}$/L$_{x}$]n$_{\rm H}$ (ergs cm$^{-3}$ s$^{-1}$).

\subsection{Comments on Planet Atmosphere Models}

Until further observations reconcile the discordant results of radial velocity
studies (Sec. 1) and establish the planet's existence beyond reasonable doubt,
detailed atmospheric modeling would be highly speculative and is not 
yet warranted. However, we summarize below some relevant factors and limitations
that will need to be confronted in any future atmospheric modeling studies.

A realistic atmospheric model requires knowledge
of the host star properties and the planet's separation
and physical properties. The separation and
relevant properties of V830 Tau itself are now reasonably well
known (Table 1), except for its EUV luminosity which must
be estimated using indirect methods (Sec. 4.3).
The suspected planet's physical properties are mostly unknown
except for a mass estimate and  orbital period.
  
The key quantity needed to calculate the
XUV heating rate  (eqs. [4], [5]) is $n_{\rm H}$($z$), 
the number density versus height $z$ in the planet's atmosphere,
where $z$ is typically defined as the distance above a fiducial
planet radius R$_{p}$ (but as already noted, R$_{p}$ is not 
known for V830 Tau b). The number density is related to the 
mass density by the usual expression
$n_{\rm H}$ =  $\rho$/($\mu$m$_{\rm H}$) where m$_{\rm H}$
is the mass of hydrogen and $\mu$ is the mean weight
per particle (amu) in the atmosphere. Not to be
overlooked is the dependence of $\mu$  on chemical
composition and state of the atmospheric gas (molecular,
atomic, ionized). Both composition and gas state  
are expected to vary with height, as is known
to occur even in the Earth's atmosphere. Chemical composition
of hot Jupiter atmospheres is generally not observationally
constrained except for a few well-studied objects like  
HD 209458 b (Charbonneau et al. 2002; Cody \& Sasselov 2002).
Thus a H-dominated atmosphere or H$+$He mixture is usually assumed,
as will probably be necessary for V830 Tau b.

Most current hot Jupiter atmosphere models are 
initialized by assuming a
base level density (or pressure) and temperature at
a reference height in the atmosphere. The base level
($z$ = 0) is often taken to be the planet's radius R$_{p}$.
The base level values are then
extrapolated upward to obtain the run of temperature (T),
pressure (P), or density with height (e.g. Murray-Clay et al. 2009; 
Yelle 2004; Salz et al. 2016). There is considerable variation 
in assumed base level values and simulated T-P profiles even for  
HD 209458 b (Fig. 9 of Vidal-Madjar et al. 2011).

As a first approximation for base level temperature the planet's  
equilibrium temperature can be used,
T$_{eq}$ = T$_{\rm eff}$[R$_{*}$/2$a$]$^{0.5}$[$f$(1 - A$_{B}$)]$^{0.25}$,
where A$_{B}$ is the planet's Bond albedo and $f$ = 1 for  even
heating of the atmosphere (Seager et al. 2000). Using the stellar
parameters for V830 Tau (Table 1) and assuming A$_{B}$ $\approx$ 0.05 - 0.1
for a hot Jupiter (Sudarsky et al. 2000) gives T$_{eq}$ $\approx$ 1200 K.
This expression accounts for the star's photospheric heating but not
XUV heating or any internal planet heating so would need to be considered a 
lower limit for V830 Tau b.

The value to be used for the planet's base density (or pressure) 
corresponding to the adopted base temperature is quite uncertain.
Some guidance is provided by theoretical T-P models of different
classes of extrasolar giant planets (e.g. Sudarsky et al. 2003).
Uncertainties in the base level density propagate to higher 
levels and thereby affect the heating rate at a given height.
Heating in the upper atmospheres of hot Jupiters is believed
to result in mass loss via a slow planetary wind which must
be modeled hydrodynamically (e.g. Yelle 2004; Murray-Clay et al. 2009;
Owen \& Jackson 2012). In the hydrodynamic picture the mass loss rate 
for a steady spherical wind is (Lamers \& Cassinelli 1999)
$\dot{M}_{p} = 4\pi r^{2} \rho_{wind}(r) v_{wind}(r)^{2}$
where $r$ is the radial distance from the planet's center,
$\rho_{wind}(r)$ is the wind's mass density, and
$v_{wind}(r)$ is the wind speed. Uncertainties in the base
level density translate into uncertainties in the wind
density and $\dot{M}_{p}$, as discussed by Salz et al. (2016).

\newpage
\section{Summary}

\begin{enumerate}
      
\item We have presented new {\em Chandra} observations of V830 Tau, a T Tauri
      star whiich may host a hot Jupiter. The star's X-ray emission is
      characterized by multi-temperature plasma viewed through low absorption and 
      is variable. The {\em Chandra} observations give an X-ray luminosity
      log L$_{x}$(0.2-8 keV) =  30.10 - 30.58 ergs s$^{-1}$ but previous 
      {\em XMM-Newton} observations reveal values as high as
      log L$_{x}$ = 30.87 ergs s$^{-1}$. 

\item The {\em Chandra}  MEG1 grating spectrum shows emission lines spanning
      a range of maximum line power temperatures from Ne IX (T$_{max}$ = 4 MK)
      to S XV (T$_{max}$ = 16 MK). No significant centroid shifts were detected
      in the brightest lines (Ne IX and Ne X). The Ne IX triplet intercombination
      ($i$) line was not detected and there is no indication of a suppressed
      forbidden-to-intercombination Ne IX line flux ratio that if present could
      signify high electron densities. These properties are consistent with
      variable X-ray emission formed in a magnetically-active coronal environment,
      as is generally found for other weak-lined non-accreting T Tauri stars.

\item Adopting log L$_{x}$(0.2-8 keV) = 30.40 ergs s$^{-1}$ as a typical 
      value for V830 Tau, the unattenuated X-ray flux at the separation
      $a$ = 0.057 au of the suspected planet is F$_{x}$(0.2-8 keV) = 2.75 $\times$ 10$^{5}$ 
      ergs cm$^{-2}$ s$^{-1}$. This is a factor of
      $\sim$10$^{6}$ - 10$^{7}$ times greater than the Sun's X-ray flux at Jupiter
      during active and quiet states. 

\item X-ray ionization and heating rates of the planet's atmosphere were computed 
      based on the star's characteristic X-ray temperature and luminosity.  
      An expression for the EUV heating rate is also given but is subject
      to uncertainty in the star's EUV luminosity, which is not directly
      measurable.  Numerical values of the X-ray and EUV heating rates depend 
      on the run of number density with height $n_{\rm H}$($z$) in the planet's
      atmosphere, which remains to be determined using hydrodynamic models. 

\item Detailed atmospheric models will be justified if the planet's existence
      is firmly established. We have identified several areas of uncertainty
      that will need to be addressed in any future models. These include 
      unknown or poorly-contrained quantities such as the star's EUV luminosity, 
      the planet's radius and  chemical composition, and base level temperature, 
      pressure, and density deep in the atmosphere.

\item Additional X-ray monitoring of V830 Tau would be useful to 
      determine if its X-ray variability is phased to the stellar
      rotation period or the planet's reported orbital period.
      
\end{enumerate}

\begin{acknowledgments}
Support for this work was provided by  {\em Chandra} award 
number GO9-20009X issued by the {\em Chandra} X-ray Center, which is operated by
the Smithsonian Astrophysical Observatory (SAO) for and on behalf of NASA.
\end{acknowledgments}

\vspace{5mm}
\facilities{{\em Chandra X-ray Observatory}}

\vspace{5mm}
\software{CIAO (Fruscione et al. 2006),
          XSPEC (Arnaud 1996)}

\clearpage

\end{document}